\documentclass{article}

\PassOptionsToPackage{numbers, compress}{natbib}

\usepackage[final]{neurips_2021}

\usepackage[utf8]{inputenc} 
\usepackage[T1]{fontenc}    
\usepackage{hyperref}       
\usepackage{url}            
\usepackage{booktabs}       
\usepackage{amsfonts}       
\usepackage{nicefrac}       
\usepackage{microtype}      
\usepackage{xcolor}         
\usepackage{graphicx}
\title{Qualitative Analysis for Human Centered AI}
\usepackage[hyphenbreaks]{breakurl}

\author{%
 Orestis Papakyriakopoulos$^{1,2}$  \\
  \texttt{orestis@princeton.edu} \\
   \And
  Elizabeth Anne Watkins$^{1,2}$  \\
   \texttt{ ew4582@princeton.edu} \\
   \And
  Amy Winecoff$^{1,2}$  \\
   \texttt{aw0934@princeton.edu} \\
   \And
  Klaudia Jaźwińska$^1$  \\
  \texttt{klaudia@princeton.edu} \\
   \And
   Tithi Chattopadhyay$^1$  \\
  \texttt{tithic@princeton.edu} \\
  \AND
  \\
  $^1$Center for Information Technology Policy\\
  Princeton University \\
  \\
  $^2$Authors contributed equally. 
}

\begin{document}

\maketitle

\begin{abstract}

Human-centered artificial intelligence (AI) posits that machine learning and AI should be developed and applied in a socially aware way. In this article, we argue that qualitative analysis (QA) can be a valuable tool in this process, supplementing, informing, and extending the possibilities of AI models. We show this by describing how QA can be integrated in the current \textit{prediction paradigm} of AI, assisting scientists in the process of selecting data, variables, and model architectures. Furthermore, we argue that QA can be a part of novel paradigms towards Human Centered AI. QA can support scientists and practitioners in practical problem solving and situated model development. It can also promote participatory design approaches, reveal understudied and emerging issues in AI systems, and assist policy making.
\end{abstract}

\section{Introduction}

From the perspective of human-centered AI (HCAI), machine learning (ML) and AI systems should be designed and applied in ways that enhance human abilities and acknowledge sociocultural context. Following from value-sensitive design, AI and ML systems should support human needs while also being fair, trustworthy, reliable, and safe \cite{aragon2022human,shneiderman2020bridging,xu2019toward,riedl2019human}. Yet, the predominant approach to ML model development often does not fully realize the goals of HCAI, at times even coming into direct conflict with them. Most data scientists and ML practitioners train models to maximize the accuracy of predicting an outcome metric or key performance indicator. Such an approach ignores the complex real-world sociotechnical systems in which models will ultimately be deployed as well as the innumerable factors that will determine how the models will be interpreted and used \cite{suchman1987plans}. Although predictive accuracy is often relevant to human needs, it is rarely the only metric that matters. In some cases, a highly accurate model may even be a poor fit for human needs. For example, algorithms for recommendation systems are still primarily assessed via accuracy metrics on historical offline datasets \cite{jannach2016recommendations}, but a model that can accurately recommend something to a user that she has already consumed is unlikely to be of much use \cite{mcnee2006being}. Thus, an approach that focuses singularly on predictive accuracy is unlikely to meet users' needs and may harm users, even if unintentionally (e.g., \cite{Hagey2021}). 

Qualitative analysis (QA) can address these issues by providing deeper insight into human experiences and perspectives that can inform the development of HCAI in at least two ways. First, QA can improve the human-centricity of the decisions data scientists and ML developers make within the predominant predictive framework. Second, QA can suggest other frameworks for developing ML and AI that can meet users' more complex needs and respect the social-embeddedness of models in users' lives. Even though HCAI involves many facets, in our discussion, we primarily focus on fairness as one illustrative example.


\section{The problem with prediction}

ML approaches aim to create models of the world that generate useful predictions \cite{breiman2001statistical, kuhn2013applied}. This \textit{prediction paradigm} \cite{gill2020prediction} is applied in numerous high-stakes contexts, such as predicting "criminality" \cite{chouldechova2017fair}. Even for ostensibly lower-stakes circumstances such as predicting the best worker for delivering a given food order \cite{griesbach2019algorithmic}, algorithmic decisions can have a significant impact on workers, such as by shaping their potential income or their work schedule \cite{dzieza_2021}. Yet a simplified process of optimizing models based on a general notion of predictive accuracy can yield two specific classes of fairness issues, when not taking social embeddedness into consideration. First, asymmetric predictive accuracy occurs when the model learns the preferences and patterns of a majority user subgroup, effectively ignoring the preferences and needs of a minority subpopulation. Second, even if the model does learn patterns from the minority subpopulation of users, if the data contains systematic social biases, models are more likely to harm marginalized users than help them by reproducing existing biases and systems of power. 

With respect to asymmetric predictive accuracy, social data systematically contain a population bias \cite{barocas-hardt-narayanan}, over-representing in the dataset socially dominant groups and individuals and under-representing minorities and isolated populations. Predictions derived from models trained on such datasets will be both less accurate and less consistent for underrepresented groups compared to the majority user subgroup. Where the minority subpopulation's data patterns deviate from the patterns of the dominant group, these deviations are effectively treated as error. From an HCAI perspective, no subpopulation treatment should be treated as an "error." Such model behavior has already been traced in computer vision \cite{buolamwini2018gender} and voice recognition  \cite{koenecke2020racial}. 

Even in instances where models do not fail to learn about underrepresented cases in the data, there is no guarantee that what the model learns about marginalized users will be beneficial to them. In such a case, preexisting biases and social structures in datasets are diffused and/or even reinforced through AI models \cite{friedman1996bias}, leading to unfair model inferences, replication of social asymmetries and historical inequalities. For example, natural language processing (NLP) models replicate biases existing in the input data leading to the generation of inferences in downstream tasks that might be racist, sexist, or replicate white-patriarchal social structures \cite{papakyriakopoulos2020bias}. Predictive policing tools can create feedback loops where police are deployed to areas that have been policed in the past, which reinforces negative stereotypes and degrades the targeted communities' quality of life \cite{lum2016predict}. Similarly, when search engines reproduce stereotypical associations people make about individuals, social groups, and specific phenomena, they further disseminate these stereotypes to new users who might learn to accept harmful associations as the algorithmic "normal"\cite{metaxa2021image}.

\section{Improving the existing paradigm with QA}

The inclusion of QA in the existing prediction paradigm can help data scientists avoid several so-called "traps" \cite{selbst2019fairness} of machine-learning development, and better equip ML developers to achieve their goals with justice, equity, and efficacy. Whereas current norms around the framing of engineering problems tend to ignore contexts and communities into which algorithms will be integrated as "solutions," the "traps" approach argues that negative outcomes await developers who fail to include context as a key consideration in ML development. Selbst et al. identify six traps. In the Framing Trap, the authors argue that ML solutions can cause harm if the humans and communities who will be using ML tools are not adequately considered as a key part of ML solution framing. In the Portability Trap, the authors points out that in the pursuit of scaled-up solutions, ML systems are often implemented into contexts for which they were not designed, and that this mismatch may create unanticipated negative outcomes. In the Formalism Trap, the authors suggest that some concepts, such as fairness, are highly contingent social ideas, not easily reduced to formal mathematical concepts, yet such reductions all too often take place in exactly this fashion, to the detriment of impacted communities. In the Ripple Effect Trap, the authors recognize that ML solutions, indeed any significant technological tool, are not situated in a vacuum, and that their implementation can provoke changes in a context which can ripple outward in surprising ways. Finally, in the Solutionism Trap, perhaps the most endemic in ML development, authors argue that technological tools may not be the best solution for a problem at all.  Qualitative analysis methods can assist in addressing all of these traps, including helping model developers pay greater attention to the context, social conditions, and human beings who will eventually use and be impacted by algorithmic systems, in the process of selecting variables, data, and model architecture. In the following, we illustrate how QA can assist ML researchers and practitioners in these processes, all the while accounting for the aforementioned ``traps'' that exist in the machine-learning development pipeline.


\begin{itemize}

\item \textbf{Variables $x$}

QA can identify both input variables that are associated with desired outcomes and input variables that can result in unintended harms, which might otherwise be overlooked if QA was not conducted. For example, many companies offer AI tools for hiring \cite{raghavan2020mitigating}. In developing these tools, a qualitative study can identify which factors might discriminate against specific social groups, allowing model developers to account for these factors in the model design process and avoid potential Framing traps existing in the specific ML application. As a result, AI tools used to assist human resource (HR) professionals would be less likely to perpetuate biases and social asymmetries from past hiring decisions \cite{schumann2020we}. Even when group specific variables are important for model robustness, researchers already provide techniques for how to correct them \cite{zhang2018mitigating}. For example, a model that learns to predict the income of a person based on historical data will have higher accuracy when it includes gender in its feature dataset, but gender should not play a role when applying a model to new predictions in a hiring setting. QA can therefore function as a guide for detecting such input variables and then allow data scientists to take appropriate steps. Furthermore, models often fail to make satisfactory predictions about phenomena, even when extensive data collection has already taken place. An example is the  Fragile Families and Child Well-being Study, which collected data about individuals' life outcomes. Multiple research groups developed different machine-learning methods, and used a dataset that was ``painstakingly collected by social scientists over 15 years,'' to predict life outcomes such as a child’s grade point average and whether a family would be home evicted\cite{salganik2020measuring}. Nevertheless, the study demonstrated that a lot of complex techniques were not adequately able to predict life outcomes. In cases such as this, QA can be a valuable tool for
investigating incidents of extreme model failure, in order to understand whether
important variables and data that is difficult to quantify were not taken into consideration during the formulation of the study (Framing trap), or even to understand the limits of predictive ML and recognize a Solutionism Trap.

\item \textbf{Data $D$}

In the case of AI reproducing unintended outcomes that are hidden in the input dataset, QA can help identify unwanted model outcomes and the social context in which they emerge. This can provide valuable knowledge that can guide data scientists in the data selection process, creating an observation space that is socially beneficial. For example, deep-learning culture in NLP and computer vision tends to include as many information sources as possible at the model-training phase, in order to improve models' accuracy. Nevertheless, this leads to problematic outcomes, as data sources such as Wikipedia include historical asymmetries and problematic structures \cite{graells2015first}. Although texts from different sources might contain the same variables (words), the associations existing in these datasets can have a significantly different impact on what a model learns. Therefore, scientists should not only reflect on the variables they collect, but also the nature of the observations that include these variables. QA can help to connect harms to specific datasets of observations, allowing data scientists to then choose appropriate data or techniques to mitigate them, in ways that avoid the Portability and Ripple Effect Traps. In this way, models can be trained on specially curated datasets that account for different issues existing in different parts of the statistical population. For example, many computer vision researchers have retracted their datasets from open usage after finding that models trained on them resulted in unintended consequences \cite{peng2021mitigating}. Furthermore, researchers offer techniques to transform datasets in specific ways in order to address specific issues of unfairness \cite{ramaswamy2021fair,salimi2019interventional}.

\item \textbf{Cost-function $L$}

QA can identify, characterize, or clarify cases of unfairness, harm, or user dissatisfaction that might have not otherwise be visible to AI designers. Informed by a better understanding of unwanted effects of an AI model, developers can account for these considerations formally in a model's cost function, developing mathematical frameworks that fulfill the goals of HCAI. For example, in the field of algorithmic fairness, researchers have developed techniques ensure a model is optimized by satisfying statistical parity among social groups \cite{dwork2012fairness,amini2019uncovering,he2020geometric}. QA can be used to identify whether a specific cost-function would be appropriate. It can also be used to detect issues of a specific cost-function: whether the mathematical model does not account for a specific instance of justice in a specific context, hence uncovering a Formalism Trap. Overall, QA can detect issues that are important to the users and stakeholders that are in the environment of an AI system, and data scientists can then account for them during model optimization. Such an approach can be beneficial in algorithmic management systems such as online labor platforms, where different stakeholders hold different conflicting needs \cite{wood2019good}. Detecting these needs by QA and taking them into consideration in algorithmic processes has the potential to yield more socially aware equilibria in these systems.

\end{itemize}

HCAI approaches should offer alternatives that allow for training and deploying ML models in ways that respect user groups and consider their needs. QA can be a tool in this process because qualitative research can help data scientists understand individuals' needs and problems as well as social conditions and context. Either before or after an AI system has been deployed, qualitative interviews with users whose perspectives are relevant to the system can inform decisions about three specific aspects of the standard model development process: variables $x$, data $D$, and the model cost function $L$. 

\section{Defining a novel paradigm with QA}

QA can be useful not only when integrated in the existing paradigm of ML development, but it can also serve as the conceptual foundation for a new socially aware paradigm that aims to create HCAI applications. In the following, we provide three core components of this new paradigm that use QA as a tool for connecting AI model development with the society.

\begin{itemize}

\item \textbf{Practical Problem Solving}

As ML models shift increasingly from theoretical, epistemic experiments to real-world applications, ML developers must recognize that their methods, too, must shift. A pragmatic approach that includes mixed methods may be better suited to applied ML research, demanding that ML developers set aside the pursuit of theoretical ideas in favor of “solving practical problems in the real world” \cite{small2011conduct}. HCAI is the recognition that AI systems will be integrated into social worlds, and bounded by real-world communities, constraints, and infrastructures. The shift to solving practical problems can be made more effective by integrating QA into scientists' toolkits, since QA comes in direct contact with the immediate environment of an AI application. QA can provide information about how ML models should be developed, taking into consideration actual issues and social conditions, in contrast to traditional ML development which takes place in sandbox settings. This information is not only useful for developing more human-centered AI models, but also for generating valuable scientific knowledge about the interaction of new technologies with individuals and social groups.

\item \textbf{Situated Model development}

The ML community itself has long called for greater participation of social scientists in development work \cite{irving2019ai,sloane2019ai}, which aligns with our own perspective that methodological tools used in the social sciences---including ethnographic, sociological, and anthropological pursuits---should be integrated into the ML development process. In particular, social science methods more generally and qualitative methods more specifically can proactively identify ethical risks of AI technologies before they can negatively impact users.  For example, facial verification in ride-hailing, as used by Uber via services provided by Microsoft, has had the real-world impact of erroneously deactivating transgender drivers, cutting them off from their livelihood \cite{melendez_2018}. As of 2021, Uber has instituted a process for transgender drivers to update their photos prior to being required to comply with their facial verification protocol. We do not know to what extent qualitative methods were a part of Uber's or Microsoft's development processes prior to this launch, or whether such harms were anticipated but simply did not have an impact on system design. In any event, these system errors were identified via post-hoc qualitative investigation by academic researchers, who have recommended that these companies work to mitigate such damages. Additional harms in this space that have been surfaced via qualitative methods include the insight that this system requires drivers to engage in risky behaviors, such as navigating dangerous traffic conditions to try to pull over to comply with the technology, and stopping frequently at night in darkened parking lots \cite{watkins}. These harms have yet to be addressed, and drivers continue to be exposed to these increased risks by the demand to comply with computer-vision technology. 

\item \textbf{Identifying issues that might be become important in the future}

QA, and the rigorous examination of how AI and ML-based tools are experienced at the ground level can also do critical "anticipation" work, helping practitioners and policy makers identify risks that may not develop immediately, but farther into the future. Many issues generated by AI technologies affect small parts of the population. But as these technologies are scaled across more and more domains, affected populations will grow. QA can identify such harms early on and assist AI developers in mitigating them in a timely manner.

One pathway lies through participatory design, which itself is a mode of situated model development. QA provides a tool by which to directly enroll impacted communities into the ML development process, and to use the expertise of these communities to identify potential future risks \cite{martin2020participatory}.\footnote{Practitioners of these strategies must ensure that they engage with and compensate participants with care and respect, and that their input is meaningfully integrated into system design \cite{sloane2020participation}.} In this way, issue identification can be more inclusive and also relate to issues that are currently faced by fewer individuals, but are likely to grow within the user population. Furthermore, QA can be used to identify  \textit{how} participatory design itself should be deployed.  Delgado et al.\cite{delgado2021stakeholder} argue that increasing stakeholder participation in AI design requires an understanding of imminent power structures in the sociotechnical context---why participation is needed and in which form, who should be involved and for what purpose. QA provides toolkits for answering these questions by interviewing different actors existing in the social context of an AI application. In this way, QA structures the problem space and generates knowledge that can be used to predict and prevent unwanted future outcomes.

QA can also contribute empirical evidence to algorithm policy. Evidence garnered via qualitative research can contribute instrumentally, by revealing concepts overlooked by policy makers \cite{merton1949role}; symbolically, by justifying a particular policy approach; and conceptually, via general enlightenment around a problem space\cite{greenhalgh2016research}. In the policy context, it is important that QA is not used to generalize findings, but rather to make compelling arguments about the population studied or provide details about how an intended deployment works in everyday environments. QA is an effective tool in policy making when used to communicate impact on certain populations.

\end{itemize}

As we've seen here, QA affords myriad opportunities to define, build, test, and govern algorithmic systems in ways that integrate community perspectives, anticipate potential harms, and provide inroads towards effective policy.  

\section{Conclusion}

Historically, data scientists and ML practitioners have not included QA as a part of their typical ML and AI modeling approaches. However, the existing paradigm for developing ML and AL models has repeatedly failed to achieve the ideals of HCAI. Using fairness concerns as motivating examples, we have argued that QA can improving the human-centricity of AI either by providing valuable insights into how the existing predictive paradigm can be improved or by suggesting entirely new paradigms. We have not, however, provided concrete or explicit direction on how these suggestions can be realized in practice. Future work should identify challenges for implementing QA in applied ML and AI development contexts, acknowledging the constraints that ML and AI development environments present.

\bibliographystyle{plainnat}
\bibliography{bibliography.bib}

\end{document}